\documentclass{aa}
\usepackage{graphicx}
\usepackage{epsfig,graphicx,rotating,portland,fancyheadings,longtable}
\usepackage{multirow}
\usepackage{txfonts}

\def\lsim{{_<\atop^{\sim}}}
\def\gsim{{_>\atop^{\sim}}}
\def\nup{$\nu_{\rm peak}^S$}

\begin{document}
\title{The seven year Swift-XRT point source catalog
  (1SWXRT) \thanks{Table 4 is only available in electronic form at
    the CDS via anonymous ftp to cdsarc.u-strasbg.fr (130.79.128.5) or
    via http://cdsweb.u-strasbg.fr/cgi-bin/qcat?J/A+A/text}}

\author{ V. D'Elia$^{1,2}$, M. Perri$^{1,2}$, S. Puccetti$^{1,2}$,
  M. Capalbi$^{1,2}$, P. Giommi$^{1}$, D. N. Burrows$^{3}$,
  S. Campana$^{4}$, G. Tagliaferri$^{4}$, G. Cusumano$^{5}$,
  P. Evans$^{6}$, N. Gehrels$^{7}$, J. Kennea$^{3}$, A. Moretti$^{4}$,
  J. A. Nousek$^{3}$, J.P. Osborne$^{6}$, P. Romano$^{5}$ and
  G. Stratta$^{1,2}$
}

\institute {
  $^{1}$ASI-Science Data Centre, Via Galileo Galilei, I-00044 Frascati, Italy;\\
  $^{2}$INAF-Osservatorio Astronomico di Roma, Via Frascati 33, I-00040 Monteporzio Catone, Italy;\\
  $^{3}$Department of Astronomy and Astrophysics, Pennsylvania State University, University Park, Pennsylvania 16802, USA; \\
$^{4}$INAF-Osservatorio Astronomico di Brera, via E. Bianchi 46, 23807 Merate (LC), Italy;\\  
$^{5}$INAF-Istituto di Astrofisica Spaziale e Fisica Cosmica di Palermo, Via U. La Malfa 153, I-90146, Palermo, Italy; \\
$^{6}$Dept. of Physics and Astronomy, University of Leicester, Leicester, LE1 7RH, UK.\\
$^{7}$NASA-Goddard Space Flight Center, Greenbelt, Maryland, 20771, USA; \\
}

  \abstract 
  {The {{\it Swift}} satellite is a multi-wavelength observatory
    specifically designed for gamma-ray burst (GRB) astronomy that is
    operational since 2004.  Swift is also a very flexible
    multi-purpose facility that supports a wide range of scientific
    fields such as active galactic nuclei, supernovae, cataclysmic
    variables, Galactic transients, active stars and comets. The Swift
    X-ray Telescope (XRT) has collected more than 150 Ms of
    observations in its first seven years of operations.}
  {The purpose of this work is to present to the scientific community
    the list of all the X-ray point sources detected in XRT imaging
    data taken in photon counting mode during the first seven years of
    Swift operations. All these point-like sources, excluding the
    Gamma-Ray Bursts (GRB), will be stored in a catalog publicly
    available (1SWXRT).}
  {We consider all the XRT observations with exposure time longer than
    $500$ s taken in the period 2005-2011.  Data were reduced and
    analyzed with standard techniques and a list of detected sources
    for each observation was produced. A careful visual inspection was
    performed to remove extended, spurious and piled-up sources. Finally,
    positions, count rates, fluxes and the corresponding uncertainties
    were computed.}
  {We have analyzed more than 35,000 XRT fields, with exposures
    ranging between $500$ s and $100$ ks, for a total exposure time of
    almost $140$ Ms. The catalog includes approximately 89,000
    entries, of which almost 85,000 are not affected by pile-up and
    are not GRBs.  Considering that many XRT fields were observed
    several times, we have a total of $\sim$ 36,000 distinct celestial
    sources. We computed count rates in three energy bands: $0.3-10$
    keV (Full, or F), $0.3-3$ keV (Soft, or S) and $2-10$ keV (Hard,
    or H). Each entry has a detection in at least one of these
    bands. In particular, we detect $\sim$ 80,000, $\sim$ 70,000 and
    $\sim 25,500$ in the F, S and H band, respectively. Count rates
    were converted into fluxes in the  $0.5-10$, $0.5-2$ and
    $2-10$ keV bands. The flux interval sampled by the detected
    sources is $7.4\times 10^{-15} - 9.1\times 10^{-11}$, $3.1\times
    10^{-15} - 1.1\times 10^{-11}$ and $1.3\times 10^{-14} - 9.1\times
    10^{-11}$ erg cm$^{-2}$ s$^{-1}$ for the F, S and H band,
    respectively. Some possible scientific uses of the catalog are
    also highlighted. }
{}

  \authorrunning {} 
\titlerunning {GRB\,100614A and 100615A: two
    peculiar dark GRBs}

\maketitle
%

\section{Introduction}

The {\it Swift} Gamma-Ray Burst Explorer (Gehrels et al. 2004) is a
NASA mission, successfully launched on 2004 Nov. 20. The hardware and
software were built by an international team involving US, United
Kingdom and Italy, with contributions from Germany and Japan. The main
scientific driver of the {\it Swift} mission is to detect gamma-ray
bursts (GRBs) in the hard X-ray band with the Burst Alert Telescope
(BAT, Barthelmy et al. 2005) and quickly follow-up their emission at
longer wavelength with the X-Ray Telescope (XRT, Burrows et al. 2005)
and Ultraviolet/Optical Telescope (UVOT, Roming et al. 2005).

Despite being specifically designed to address GRB science topics,
{\it Swift} is also an effective multi-purpose multi-frequency
observatory. The {\it Swift} team expertise in following up GRBs has
grown during the satellite operations, leading to an evolution of the
observing time share between GRBs and other cosmic sources.  At the
beginning of the satellite operations (up to 2006, Romano 2012),
approximately $56\%$ of the {\it Swift} observing time was dedicated
to GRB observations, and $\sim 26\%$ divided up between target of
opportunity (ToO, $\sim 8\%$) and ''fill-in'' observations (that are
short exposures of a variety of X-ray sources taken when {\it Swift}
was not engaged in GRB science, $\sim 18\%$). After $2006$, it became
evident that there was no need to follow up all GRBs for a very long
time. Thus, without losing too much relevant scientific information,
in $2010$ the GRB dedicated time dropped to $\sim 27 \%$ while $\sim
29 \%$ and $\sim 26 \%$ was allocated to ToO and Fill-in observations,
respectively. In the remaining $\sim 18\%$ of the time the satellite
flies through the South Atlantic Anomaly or is devoted to calibration
issues.

The {\it Swift} mission is currently producing data at a pace of about $\sim 500$
observations per month, contributing to most areas of astronomy. 
Apart from GRBs, the {\it Swift} instruments are
observing extragalactic targets, such as active galactic nuclei,
clusters of galaxies, nearby galaxies, and Galactic sources, such as
binaries, microquasars, pulsars, and all Galactic variable sources in
general.

The {\it Swift}-XRT utilizes a mirror set built for JET-X and an
XMM/EPIC MOS CCD detector to provide a sensitive broad-band (0.2-10
keV) X-ray imager with effective area of $> 120$ cm$^2$ at $1.5$ keV,
field of view of $23.6 \times 23.6$ arcmin, and angular resolution of
$18$ arcsec. The detection sensitivity is $2\times10^{-14}$ erg
cm$^{-2}$ s$^{-1}$ in $10^4$ s. The instrument can work in three
different modes (Hill et al. 2004): Photodiode (PD), Windowed-Timing
(WT) and Photon-Counting (PC) modes. Due to a micrometeorite hit on
May 27 2005, the PD mode has been disabled because of the very high
background rate from hot pixels which cannot be avoided during
read-out in this mode (Abbey et al. 2006). While the first two modes
are built to produce a high time resolution at the expense of losing
all (PD) or part (WT) of the spatial information, the latter one
retains full imaging resolution and will be the only mode exploited
here.

This paper presents the 1SWXRT catalog, which consists of all the
point-like sources detected by the XRT in its first seven years of
operations ($2005-2011$).  Updated versions of the catalog, containing
observations performed from 2012, are foreseen on timescales of about
two years. Similar catalogs have already been produced for the first
eight years of {\it Chandra} operations (Evans et al. 2010) and for
the first seven years of XMM-Newton operations (Watson et
al. 2009). The reduction and analysis method is very similar to that
adopted for the production of the ``{\it Swift} Serendipitous Survey
in deep XRT GRB fields'' (Puccetti et al. 2011), which comprises a
list of sources detected in all {\it Swift}-XRT GRB fields with
exposure times longer than $10$ ks, observed by {\it Swift} between
$2004$ and $2008$.  Our goal is complementary to that of Puccetti et
al. (2011). Instead of summing all the observations related to the
same field, we keep them separated, in order to build a catalog which
retains information about the variability of our sources. In addition,
we analyzed {\it all} the XRT observations, and not only the fields
centered on GRBs. A future work (Evans et al., in prep) will consider
all XRT fields, combined where the same field is observed multiple
times. A first catalog of extended sources has been published
  (Tundo et al. 2012), and further updates are in preparation. Our
paper is organized as follows. In Sect. 2 we briefly describe our
catalog. In Sect. 3 and 4 we present the data reduction and analysis
method, respectively. In Sect. 5 we briefly discuss the scientific
issues that can be tackled using our catalog. Finally, in Sect. 6 we
draw our conclusions.

\begin{table}[ht]
\begin{center}
  \caption{Observations and exposure times in $2005-2011^{1,2}$} 
\smallskip
\tiny
\begin{tabular}{|l|ccccccc|c|}
  \hline               
      &$2005$  & $2006$   & $2007$      & $2008$   & $2009$   & $2010$  &$2011$ & Tot \\
  \hline
  \hline
  Jan & $140$  & $211$    & $313$       & $414$    & $418$    & $489$   & $683$ & $2668$   \\
      & $0.64$ & $1.89$   & $1.85$      & $1.83$   & $1.65$   & $1.65$  & $1.85$& $11.36$  \\
\hline
  Feb & $148$  & $192$    & $295$       & $408$    & $487$    & $505$   & $646$ & $2681$   \\
      & $0.71$ & $1.47$   & $1.62$      & $1.67$   & $1.57$   & $1.47$  & $1.56$& $10.07$  \\
\hline
  Mar & $274$  & $259$    & $411$       & $401$    & $476$    & $507$   & $703$ & $3031$   \\
      & $0.96$ & $1.66$   & $1.88$      & $1.80$   & $1.68$   & $1.58$  & $1.75$& $11.31$  \\
\hline
  Apr & $164$  & $241$    & $337$       & $391$    & $372$    & $617$    & $572$ & $2694$  \\
      & $1.31$ & $1.55$   & $1.73$      & $1.73$   & $1.73$   & $1.62$  & $1.49$ & $11.16$ \\
\hline
  May & $169$  & $372$    & $418$       & $435$    & $474$    & $631$   & $640$ &  $3139$  \\
      & $1.41$ & $1.94$   &$1.76$       & $1.72$   & $1.78$   & $1.82$  & $1.68$&  $12.11$ \\
\hline
  Jun & $183$  & $244$    & $379$       & $438$    & $480$    & $589$   & $712$ &  $3025$  \\ 
      & $1.45$ & $1.63$   & $1.67$      & $1.70$   & $1.74$   & $1.79$  & $1.73$&  $11.70$ \\
\hline
  Jul & $311$  & $283$    & $472$       & $437$    & $401$    & $588$   & $667$ &  $3159$  \\
      & $1.81$ & $1.78$   & $1.73$      & $1.58$   & $1.58$   & $1.82$  & $1.72$&  $12.02$ \\
\hline
  Aug & $217$  & $301$    & $221$       & $415$    & $494$    & $543$   & $642$ &  $2833$  \\
      & $1.72$ & $1.81$   & $0.89$      & $1.62$   & $1.88$   & $1.83$  & $1.55$&  $11.30$ \\
\hline
  Sep & $221$  & $244$    & $266$       & $346$    & $449$    & $530$   & $598$ &  $2654$  \\
      & $1.78$ & $1.64$   & $1.60$      & $1.48$   & $1.64$   & $1.70$  & $1.49$&  $11.33$ \\
\hline
  Oct & $230$  & $273$    & $283$       & $417$    & $433$    & $644$   & $645$ &  $2925$  \\
      & $1.77$ & $1.72$   & $1.67$      & $1.68$   & $1.67$   & $1.74$  & $1.64$&  $11.89$ \\
\hline
  Nov & $232$  & $235$    & $381$       & $336$    & $467$    & $645$   & $754$ &  $3050$  \\
      & $1.81$ & $1.83$   & $1.76$      & $1.79$   & $1.71$   & $1.80$  & $1.67$&  $12.37$ \\
\hline
  Dec & $212$  & $272$    & $407$       & $404$    & $483$    & $681$   & $693$ &  $3152$  \\
      & $1.85$ & $1.79$   & $1.83$      & $1.77$   &$1.79$    & $1.81$  &$1.73$ &  $12.57$ \\
  \hline
\hline
  Tot & $2501$ & $3127$   & $4183$      & $4842$   & $5434$    & $6969$ & $7955$&  $35011$ \\
      & $17.21$& $20.71$  & $20.02$     &$20.38$   & $20.43$   & $20.64$& $19.85$& $139.2$ \\
\hline
  \hline
\end{tabular}
\end{center}
$^1$ The first number of each entry shows the total observations per month. 
$^2$ The second number of each entry shows the total exposure time per month, in units of Ms.  
\end{table}

\begin{figure}
\centering

\includegraphics[angle=90,width=9.4cm]{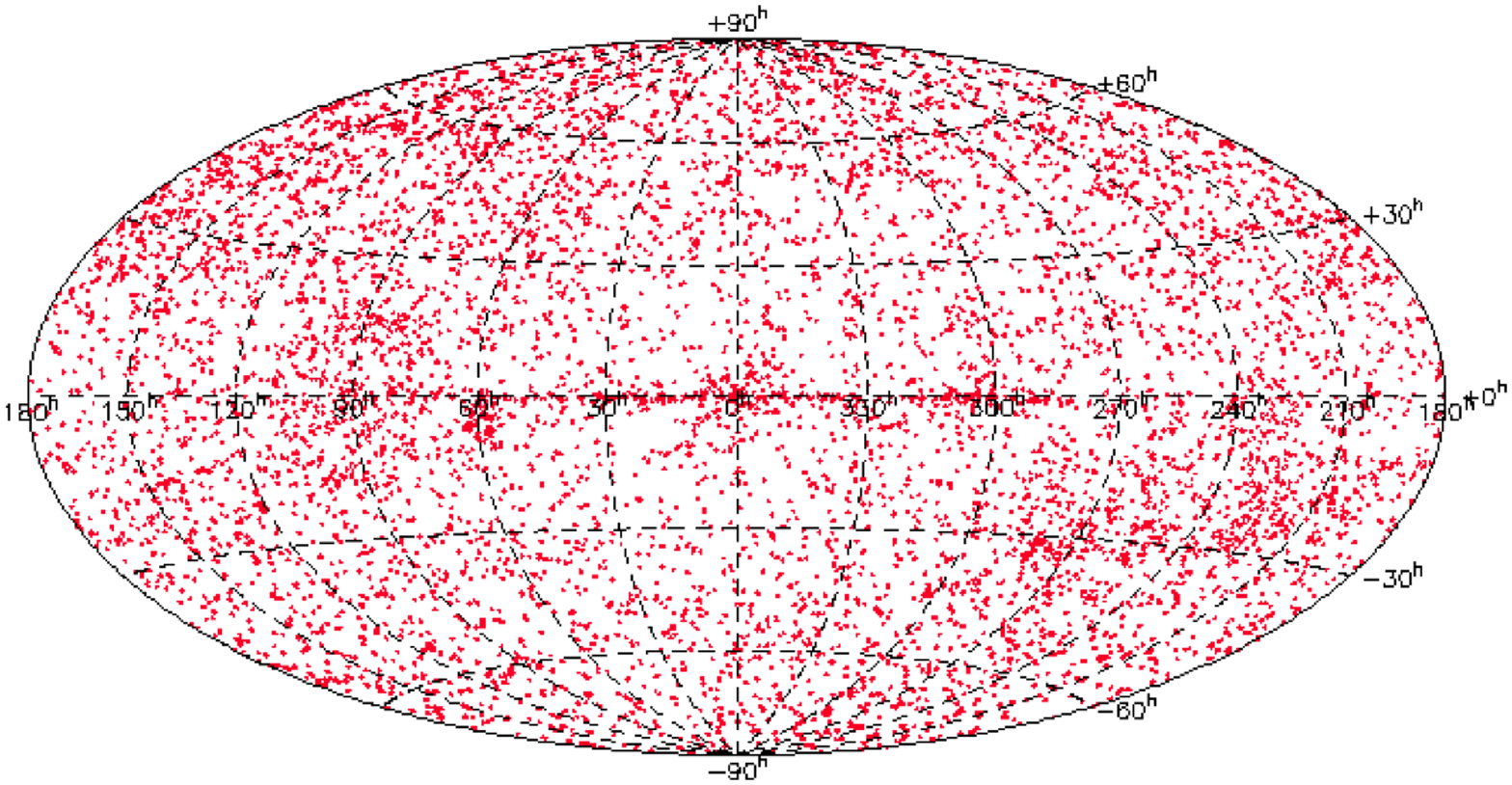}
\caption{The Aitoff projection in Galactic coordinates of the {\it
    Swift}-XRT fields analyzed in this paper.}
\end{figure}

\section{The seven year {\it Swift}-XRT point source catalog}

The seven year {\it Swift}-XRT point source catalog (1SWXRT) is built
using all the observations performed by {\it Swift}-XRT in PC
observing mode between $2005$ and $2011$. We consider in our analysis
all the XRT fields, including ``safe pointings'', that are sky
positions used by the satellite as safe positions in case there are
troubles during the slew from one target to another. The only
constraint for a field to be analyzed is on the exposure time, which
is required to be longer than $500$ s. Different observations are not
merged, but analyzed separately, thus retaining information about the
variability of the catalogued sources.  Here and in the following we
define as ``observation'' the total exposure time per target for a given
day, identified by an unambiguous sequence number.

The total number of observations considered is $35,011$, for an overall
exposure time of $\sim 140$ Ms. Fig. 1 shows the Aitoff projection in
Galactic coordinates of these XRT observations. Table 1 collects the
number of observations per month in the period $2005-2011$, together
with the corresponding exposure time. Fig. 2 shows the number of
observations as a function of the month in which they were
performed. It is interesting to note how this number increased with
time, reflecting the evolution of the {\it Swift}
observing policy. Fig. 3 displays the number of observations grouped
in bins of exposure time ($500$ s binning). Most of the observations
have short exposures. In fact, $\sim 18\%$ have $t_{exp}< 1$ ks and
$\sim 77\%$ have $t_{exp} < 5$ ks. Only $7\%$ of the observations have
an exposure time $> 10$ ks, which are mostly (but not exclusively)
fields associated with GRBs. A bump at about $10$ ks is evident in
Fig. 3. This happens because GRBs are typically observed for $10$ ks
per day, so that a lot of observations have that exposure duration.

Many of the $\sim$ 35,000 fields analyzed are repeated pointings
centered on the same sky position. To estimate the total sky coverage
of our data set, when fields were observed more than once, we
considered only the deepest exposure. This leaves us with 8,644
distinct fields, whose geometrical sky coverage as a function of the
exposure time is plotted in Fig. 4. The cumulative sky coverage of all
our distinct fields, which by definition have exposure times
$t_{exp}>500s$ is $1300$ square degrees. A full list of the
observations analyzed in this work is available online at the ASI
Science Data Centre (ASDC) website www.asdc.asi.it.

In the next sections we will discuss how these raw observations have
been reduced and analyzed, to detect and gather information on the
sources that make up the XRT catalog.

\begin{figure}
\centering

\includegraphics[angle=-0,width=9.4cm]{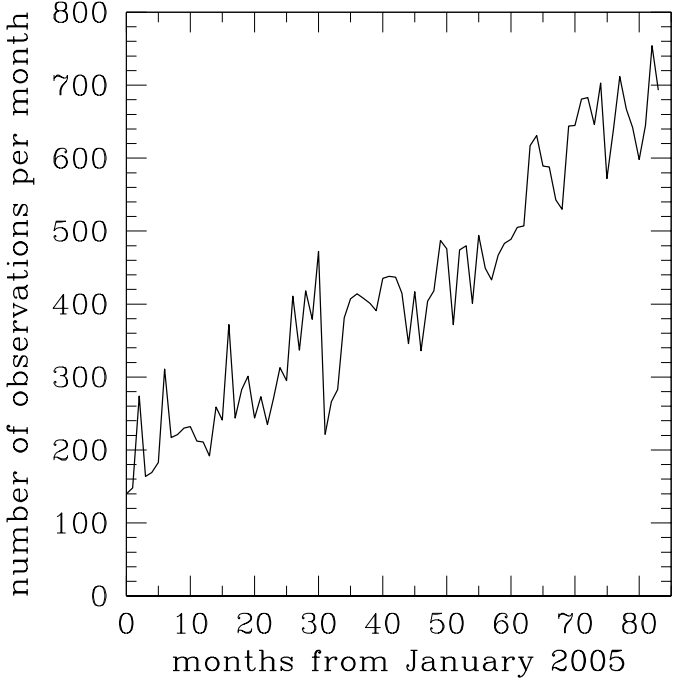}
\caption{The number of {\it Swift}-XRT observations with exposure time
  longer than 500 s acquired every month from January 1$^{st}$ 2005.}
\end{figure}

\begin{figure}
\centering

\includegraphics[angle=-0,width=9.4cm]{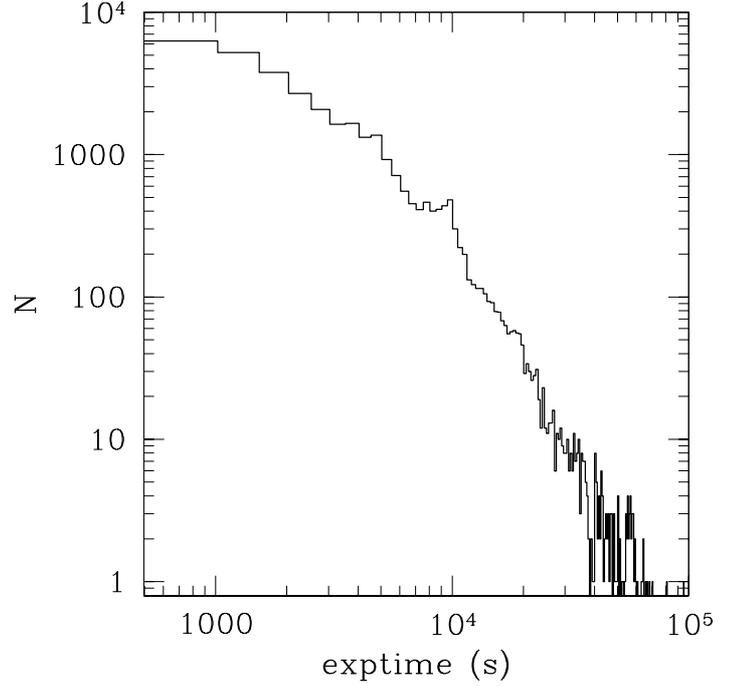}
\caption{The exposure time distribution for the XRT sources analyzed in
  this work. The time bin is $500$ s.}
\end{figure}

\begin{figure}
\centering

\includegraphics[angle=-90,width=9.1cm]{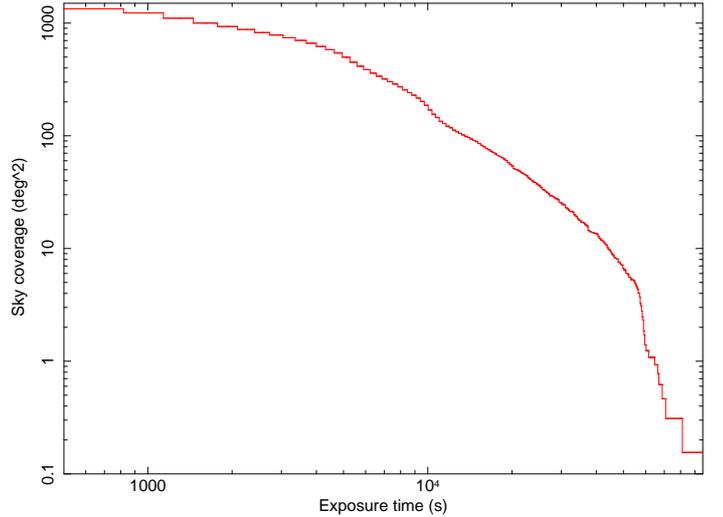}
\caption{The sky coverage of the {\it Swift}-XRT fields as a function of
  the exposure time (cumulative distribution).}
\end{figure}

\section{XRT data reduction}

The XRT data were processed using the XRTDAS software (v. 2.7.0, Capalbi et
al. 2005) developed at the ASI Science Data Centre and included in the
HEAsoft package (v. 6.11) distributed by HEASARC. For each observation of
the sample, calibrated and cleaned PC mode event files were produced
with the {\it xrtpipeline} task. In addition to the screening criteria
used by the standard pipeline processing, we applied two further, more
restrictive screening criteria to the data, in order to improve the
signal to noise ratio of the faintest, background dominated,
serendipitous sources.

\begin{figure}
\centering

\includegraphics[angle=-0,width=9.4cm]{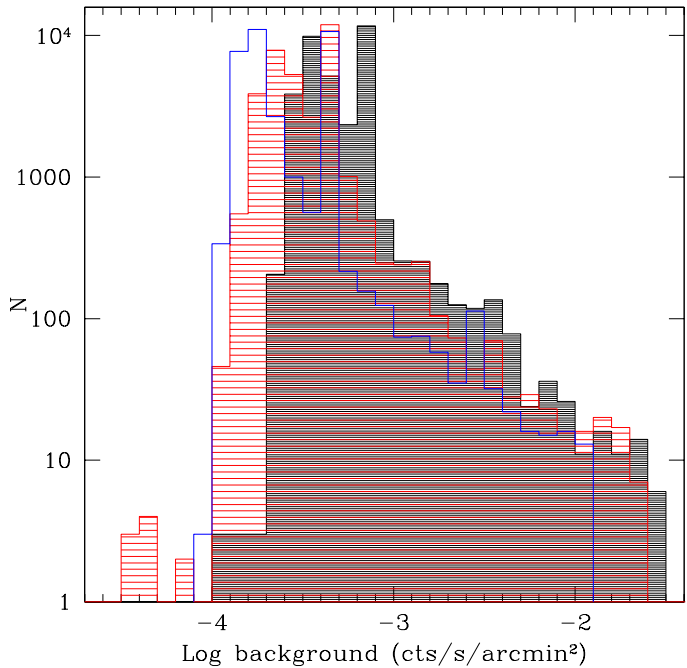}
\caption{The distribution of the mean background counts/sec/arcmin$^2$
  for our XRT observations, in the F band (black-shaded histogram), S
  band (red-shaded histogram) and H band (blue histogram).}
\end{figure}


First, we selected only time intervals with CCD temperature less than
-$50$ $^{\rm o}$C (instead of the standard limit of -$47$ $^{\rm o}$C)
since the contamination by dark current and hot pixels, which increase
the low energy background, is strongly temperature dependent. Second,
background spikes can occur in some cases, when the angle between the
pointing direction of the satellite and the bright Earth limb is
low. In order to eliminate this so-called bright Earth effect, due to
the scattered optical light that usually occurs towards the beginning
or the end of each orbit, we monitored the count rate in four regions
of $70 \times 350$ physical pixels, located along the four sides of
the CCD. Then, through the {\it xselect} package, we excluded time
intervals when the count rate is greater than $40$ counts/s.
This is enough to remove the bright Earth contamination from most (but
not all, see next section) of the XRT observations.

We produced exposure maps of the individual observations, using the
task {\it xrtexpomap}. Exposure maps were produced at three energies:
1.0 keV, 4.5 keV, and 1.5 keV. These correspond to the mean values for
a power-law spectrum of photon index $\Gamma= 1.8$ (see Sec. 4.3)
weighted by the XRT efficiency over the three energy ranges considered
here: $0.3-3$ keV (soft band S), $2-10$ keV (hard band H), $0.3-10$
keV (full band F). For each observation we also produced a background
map, using XIMAGE, by eliminating the detected sources and calculating
the mean background in box cells of size $32\times32$ pixels.

Fig. 5 shows the distribution of the mean background
counts/s/arcmin$^2$ in the F, S and H energy bands. The median values
of background and their interquartile ranges are $0.45^{+0.25}_{-0.10}$
counts/ks/arcmin$^2$, $0.31^{+0.09}_{-0.08}$ counts/ks/arcmin$^2$ and
$0.19^{+0.31}_{-0.03}$ counts/ks/arcmin$^2$ for the F, S and H band,
respectively. These median values correspond to a level of $1.1$,
$0.77$ and $0.47$ counts in the F, S, and H band, respectively, over a
typical source detection cell (see Sec. 4) and an exposure of 100 ks,
which is the highest exposure time for all our observations.

\section{Data analysis}

\subsection{Detection and filtering procedure}

The point source catalog was produced by running the detection
algorithm {\it detect}, a tool of the XIMAGE package version 4.4.1
\footnote{http://heasarc.gsfc.nasa.gov/docs/xanadu/ximage/
  ximage.html}.  {\it Detect} locates the point sources using a
sliding-cell method. The average background intensity is estimated in
several small square boxes uniformly located within the image. The
position and intensity of each detected source are calculated in a box
whose size maximizes the signal-to-noise ratio. The net counts are
corrected for dead times and vignetting using the input exposure maps,
and for the fraction of source counts that fall outside the box where
the net counts are estimated, using the PSF calibration. Count rate
statistical and systematic uncertainties are added quadratically.
{\it Detect} was set to work in bright mode, which is recommended for
crowded fields and fields containing bright sources, since it can
reconstruct the centroids of very nearby sources (see the XIMAGE
help). While producing the deep {\it Swift}-XRT catalog, Puccetti et
al. (2011) found that background is well evaluated for all exposure
times using a box size of $32\times32$ original detector pixels, and
that the optimized size of the search cell that minimizes source
confusion, is $4\times4$ original detector pixels.  The background
adopted by XIMAGE for each observation is an average of the background
evaluated in all the $32 \times 32$ individual cells. We adopted these
cell sizes and background estimation method too, and we also set the
signal-to-noise acceptance threshold to 2.5. We produced a catalog
using a corresponding Poisson probability threshold of
$4\times10^{-4}$. We applied {\it detect} on the XRT image using the
original pixel size, and in the three energy bands: F, S and H (see
Sec. 3).

The catalog was cleaned from spurious and extended sources by visual
inspection of all the observations. Spurious sources arise on the the
wings of the PSF of extremely bright sources, or near the edges of the
XRT CCD (where the exposure map drastically drops out), or as
fluctuations on extended sources and in some cases as residual bright
Earth contamination not completely eliminated by our screening
criteria. To deal with this last source of spurious detections, we
run the $detect$ algorithm on the observations affected by bright
Earth, lowering the count rate threshold value on the corners of the
detector, as defined in Sect 3. In a few cases, to avoid lowering  the
threshold excessively, and thus exclude too many time intervals from the
analysis, we decided to manually remove the spurious sources
associated with residual bright Earth contamination
even after the adopted cleaning criteria described in Sect. 3. About $
200$ observations out of the entire sample of $\sim$ 35,000 ($\sim
0.6\%$) needed a manual removal of spurious sources induced by bright
Earth background. Extended sources have also been eliminated from the
final point-like catalog, because {\it detect} is not optimized to
detect this type of sources, not being calibrated to correct for the
background and PSF loss in case of extended sources. In order to clean
the catalog from extended sources, we compare their brightness profile
with the XRT PSF at the source position on the detector, using
XIMAGE. In total, $\sim 3,700$ observations needed a manual removal of
spurious and/or extended sources, which is $\sim 10\%$ of the total
fields analyzed.

\subsection{Source statistics}


The above procedure resulted in $89,053$ point-like objects detected
in at least one of the three bands. Of these, $1,947$ are affected by
pile-up, i.e., feature more than $0.6$ counts in the full band, while
$2,166$ are GRBs, which will not appear in this catalog. After
removing GRBs and piled-up sources, we are left with $84,992$ entries,
which define a ``good'' sample.

As explained before, not all these detections represent distinct
sources, since observations of some fields are repeated many times. To
obtain an estimate of the number of independent celestial sources, we
compress our catalog over a radius of $12$ arcsec. In other words, all
entries within $12$ arcsec of each other are counted once. The choice
of the compressing radius is not straightforward. In fact, too large a
radius would lead to the compression of sources that are really
different, while too small a value would result in counting the same
source more than once, as it could have a slightly different position
in different observations due to statistical and systematic
uncertainties. We tried different compressing radii, and we noted that
the number of compressed sources increases slightly while reducing the
radius up to $12$ arcsec, while this increment is huge with a further
reduction of the compressing parameter. This means that below $12$
arcsec we are beginning to count the same sources more than once. This
number is close to twice the typical uncertainty of the weakest sources in
the XRT fields, which is roughly of 6-7 arcsec. The
estimated number of independent celestial sources obtained in this way
is $\sim$ 36,000. In this section, however, we will consider every one
of the 84,992 entries of the catalog, because of the
observation-by-observation analysis we decided to perform to build our
database.

\begin{table}[ht]
\begin{center}
  \caption{Number of sources detected in each band and any combination of them.} 
{\footnotesize \smallskip
\begin{tabular}{|l|ccc|}
\hline               
band &F      & S   & H         \\
N    &$80123$    &$70018$  & $25437$       \\
\hline   
band &F+S+H  & F+S   & F+H     \\
N    &$22016$  &$65826$   & $24751$    \\
\hline   
band &F only &S only & H only  \\
N    &$11562$&$4183$&$677$  \\
\hline

\end{tabular}
}
\end{center}

\end{table}

Table 2 shows the detections in each of the three bands and in all
possible combinations of them. In particular, 80,123 sources are
detected in the F band, 70,018, in the S band and 25,437 in the H
band.  Fig. 6 plots the histogram of the number of sources detected
per field. Most of the observations present few sources, with $\sim
51\%$ of the fields having just one or no detections and less than
$5\%$ showing more than $10$ sources. This is a consequence of the
features of our sample, composed by many observations with a low
exposure time.

\begin{figure}
\centering

\includegraphics[angle=-0,width=9.4cm]{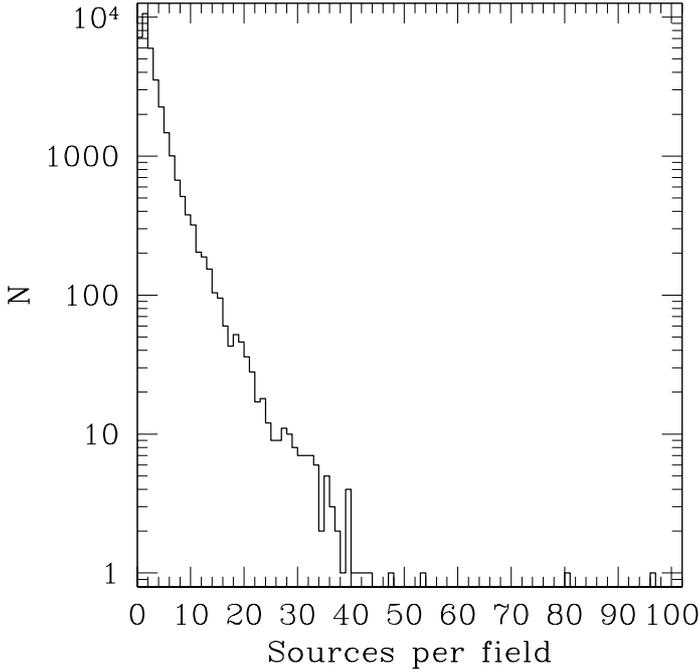}
\caption{The distribution of the detected sources per field. Each
  source in this plot is detected in at least one of the three bands.
  Fields with more than $40$ sources have high exposure times.}
\end{figure}

\subsection{Count rates and fluxes}

As explained in section 4.1, the count rates are estimated through the
{\it detect} algorithm in the F, S, H bands and corrected using proper
exposure maps (i.e., taking into account bad columns and vignetting)
and PSF. To assess the reliability of the count rates evaluated with
{\it detect}, Puccetti et al. (2011) selected a sample of 20 sources
at different off-axis angles, and compared the {\it detect} results
with that obtained by extracting the source spectra in a region of
$20$ arcsec. The average ratio between the count rates estimated using
the two methods resulted to be $1.1\pm0.2$, confirming the reliability
of our method. Fig. 7 shows the distribution of the count rates in the
three energy bands. The median values of the count rates are
$3.86\times10^{-3}$ , $3.85\times10^{-3}$ and $6.89\times10^{-3}$ cts
s$^{-1}$ in the F, S and H band, respectively. The faintest objects
have been detected in the longest exposure time observations. The
lowest count rate values estimated are $\sim 2.1\times 10^{-4}$, $\sim
1.8\times 10^{-4}$ and $\sim 1.5\times 10^{-4}$ cts s$^{-1}$ in the F,
S and H band, respectively.

\begin{figure}
\centering

\includegraphics[angle=-0,width=9.4cm]{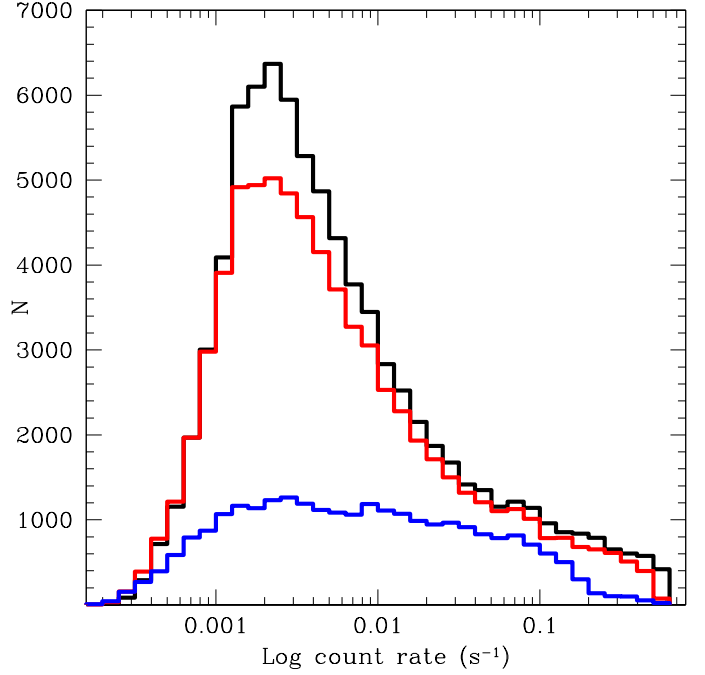}
\caption{The distribution of the count rates in the full (black), soft
  (red) and hard (blue) bands.}
\end{figure}

Count rates in the F, S and H bands were converted to $0.5-10$,
$0.5-2$ and $2-10$ keV observed fluxes, respectively. We adopted these
flux bands to be consistent with previously published works (e.g.,
Watson et al. 2009; Evans et al. 2010). The conversion was made under
the assumption that the spectral shape of each source is described by
an absorbed power-law.  The Hydrogen column density (N$_H$) in the
direction of our target is assumed to be the Galactic one, while the
photon spectral index $\Gamma$ has been estimated through the hardness
ratio\footnote{We adopt the standard notation $f(E) \propto
  E^{-\beta}$, $f(E)$ being the flux as a function of the energy. The
  photon spectral index is defined by $\Gamma \equiv \beta +1$. }. The
latter quantity is defined, for each source, as
$HR=(c_H-c_S)/(c_H+c_S)$, $c_S$ and $c_H$ being the count rates in the
S and H band, respectively. Fig. 8 plots the hardness ratio
distribution of our sources and their spectral indices. The median
value of the hardness ratio is $HR_{M}=-0.38$, while the distribution
peaks at $HR_{P}=-0.50$. However, $HR$ can be evaluated only for
objects with a detection in both the S and H bands, which are 21,097
out of a total of 84,992, i.e., $\sim 25\%$ of our sample (see Table
2). For sources which miss the detection in one of these two bands,
the $\Gamma$ slope must be chosen somehow. One way would be to compute
the average or the median of all the $\Gamma$ values of our
sources. However, this is not the best strategy, because $\Gamma$
strongly depends on the source type, and our sample is highly
heterogeneous. Thus, we decide to fix the photon index of the sources
with a missing S or H count rate to $\Gamma \equiv 1.8$, following
Puccetti et al. (2011). In fact, they computed the most probable
hardness ratio value ($HR=-0.5$) in a subsample of their catalog
comprising all the high Galactic-latitude ($|b|>20\deg$) sources.
This HR value, combined with the median of the Galactic Hydrogen
column density ($N_H=3.3\times 10^{20}$ cm$^{-2}$, Kalberla et
al. 2005), corresponds to $\Gamma = 1.8$. This choice should provide a
reliable flux estimate for our extragalactic sources, which constitute
most of our catalog. However, the reader must be aware that the flux
computed this way may represent just a rough estimate for other type
of sources (see also next sub-section for a more detailed description
about the flux uncertainties). The faintest fluxes sampled by our
survey belong to the sources detected in the deepest observations. In
detail, we find that the flux interval sampled by the detected sources
is in the range $7.4\times 10^{-15} - 9.1\times 10^{-11}$, $3.1\times
10^{-15} - 1.1\times 10^{-11}$ and $1.3\times 10^{-14} - 9.1\times
10^{-11}$ erg cm$^{-2}$ s$^{-1}$ for the F, S and H band, respectively.

\begin{figure}
\centering

\includegraphics[angle=-0,width=9.4cm]{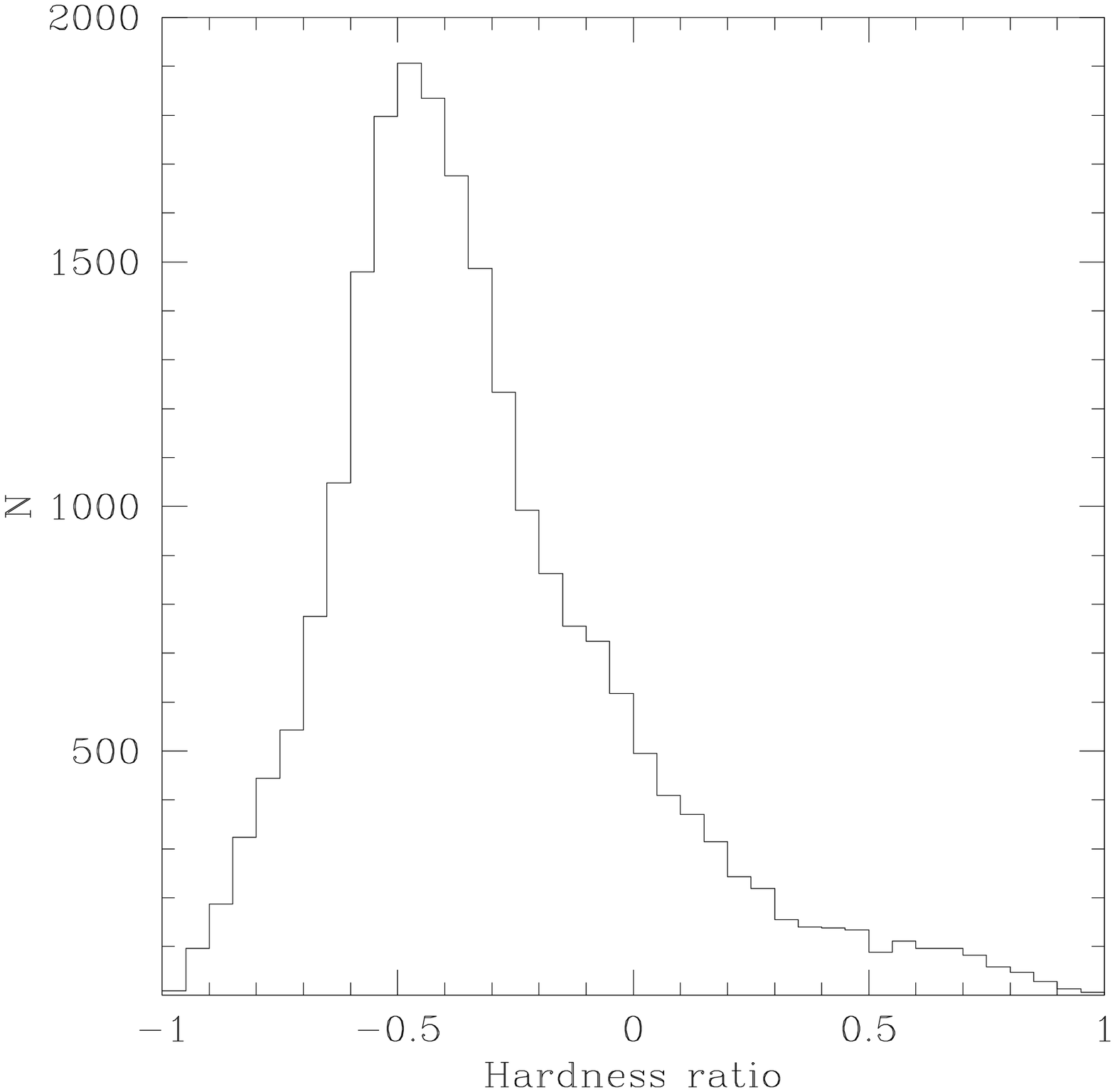}
\includegraphics[angle=-0,width=9.4cm]{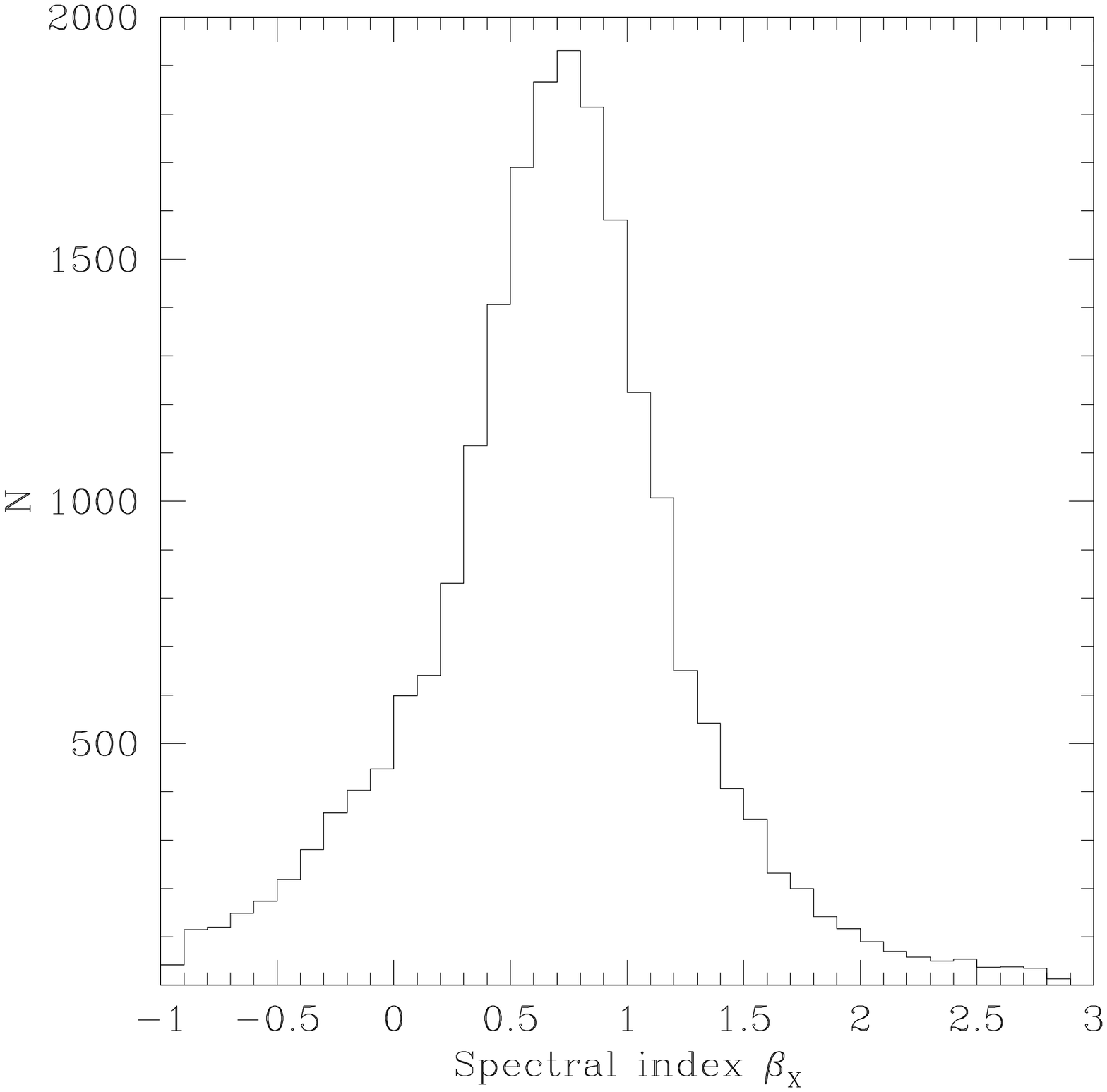}
\caption{Top panel: the hardness ratio distribution of our
  sample. Bottom panel: the distribution of the X-ray spectral indices
  for our sample. These plots include only sources with a detection in
  both the S and H bands.}
\end{figure}

We provide $90\%$ count rate and flux upper limits every time a
source is not detected in one or two of the considered bands. The
$90\%$ count upper limit for a given background is defined as the
number of counts necessary to be interpreted as a background
fluctuation with a probability of $10\%$ or less, according to a
Poissonian distribution. In other words, if the background of our
field is $B$, we are searching the upper limit $X$ for which

$$  P_{Poisson} \equiv e^{-(X+B)}\sum_{i=0}^{M}{(X+B)\over i!}  \leq 0.1, \eqno (1) $$

\noindent where $M$ is the number of counts measured at the position
of each source in a region of $16.5$ arcsec radius, which corresponds
to a fraction of the point spread function of $\sim 68\%$. Eq. (1)
does not take into account possible background fluctuations that may
arise close to the considered source. The correction factor has been
evaluated by Puccetti et al. (2011) following the recipe in Bevington
and Robinson (1992). They found that the factor $1.282\times\sigma$
(with $\sigma(B) = \sqrt B$ describing the Poissonian background
fluctuations) must be added to the count upper limits. The count rate
upper limits are finally evaluated from these counts (which are
corrected for the non-included PSF fraction of the cell), by dividing
them for the net exposure, which takes into account the vignetting at
the source position. Flux upper limits are computed from count rate
upper limits, adopting the appropriate N$_H$ and assuming $\Gamma =
1.8$, as explained before.

\begin{table*}[t]
\caption{Source parameters in the catalog.}
\begin{center}
\begin{tabular}{|rrl|}
  \hline
  Column & Parameter & Description\\
  \hline
  1 & NAME & source name: prefix 1SWXRT, following the standard IAU convention\\
  2 & TARGET NAME & XRT field name \\
  3 & RA & {\it Swift}-XRT Right Ascension in hms in the J2000  coordinate system\\
  4 & DEC & {\it Swift}-XRT Declination in hms in the J2000  coordinate system\\
  5 & SEQUENCE & {\it Swift}-XRT observation number\\
  6 & START DATE & Start time of the field observations in year-month-day h:m:s\\
  7 & END DATE & End time of the field observations in year-month-day h:m:s\\
  \hline
  8 & F\_RATE       & 0.3--10~keV count rate or 90\% upper limit in counts/sec\\
  9& F\_RATE\_ERR  & 1$\sigma$ 0.3--10~keV count rate error in counts/sec, in case of upper limits is set to -99\\
  10 & F\_FLUX       & 0.5--10~keV Flux or 90\% in erg~cm$^{-2}$s$^{-1}$  \\
  11 & F\_FLUX\_ERR  & 1$\sigma$ 0.5--10~keV Flux error in erg~cm$^{-2}$s$^{-1}$, in case of upper limits is set to -99\\
  12 & F\_PROB  & 0.3--10~keV detection probability\\
  13 & F\_SNR       & 0.3--10~keV S/N \\
  \hline
  14& S\_RATE       & 0.3--3~keV count rate or 90\% upper limit in counts/sec\\
  15& S\_RATE\_ERR  & 1$\sigma$ 0.3--3~keV count rate error counts/sec, in case of upper limits is set to -99\\\
  16& S\_FLUX       & 0.5--2~keV Flux or 90\% upper limit in erg~cm$^{-2}$s$^{-1}$\\
  17& S\_FLUX\_ERR  & 1$\sigma$ 0.5--2~keV Flux error in erg~cm$^{-2}$s$^{-1}$, in case of upper limits is set to -99\ \\
  18 & S\_PROB  & 0.3--13~keV detection probability\\
  19 & S\_SNR       & 0.3--3~keV S/N \\
  \hline
  20& H\_RATE       & 2--10~keV count rate or 90\% upper limit in counts/sec\\
  21& H\_RATE\_ERR  & 1$\sigma$ 2--10~keV count rate in counts/sec, in case of upper limits is set to -99\\
  22& H\_FLUX       & 2--10~keV Flux or 90\% upper limit in erg~cm$^{-2}$s$^{-1}$\\
  23& H\_FLUX\_ERR  & 1$\sigma$ 2--10~keV Flux error in erg~cm$^{-2}$s$^{-1}$, in case of upper limits is set to -99\\
  24 & H\_PROB       & 2--10~keV detection probability \\
  25 & H\_SNR  & 2--10~keV S/N\\
  \hline
  26 & EXPOSURE & Total on-time in sec\\
  27 & HR  & hardness ratio = (h\_rate-s\_rate)/ (h\_rate+ s\_rate) (set to 99 if soft or hard counts are missing) \\
  28 & HR\_ERR & 1$\sigma$  hardness ratio error evaluated with the error propagation formula (see e.g. Bevington \& Robinson 1992)\\
  29 & POS\_ERR & Positional error at 68\% confidence level in arcsec\\
  30 & N$_H$ & Galactic hydrogen column density in cm$^{-2}$\\
  31 & BETA & Energy spectral index (set to 99 if soft or hard counts are missing)\\
  32 & BETA\_ERR & Energy spectral index error \\
  \hline
\end{tabular}
\end{center}
\label{cata}
\end{table*}

\begin{table*}[t]
\tiny
\caption{1SWXRT catalog template}
\begin{center}
\begin{tabular}{|ccccccc|}
  \hline
  \hline
          NAME          &        TARGET NAME    &        RA        &       DEC         &     SEQUENCE &    START DATE         &         END DATE       \\
  \hline                                                                                              
  1SWXRT J074015.8-885016  &  1RXSJ073856.5-88404  &   115.066029     &  -88.8379703     &  00036747001 &  2007-12-08 01:27:33  &  2007-12-08 23:59:57  \\
  1SWXRT J073859.3-884038  &  1RXSJ073856.5-88404  &   114.747213     &  -88.6774522     &  00036747001 &  2007-12-08 01:27:33  &  2007-12-08 23:59:57  \\
  1SWXRT J073924.2-884036  &  1RXSJ073856.5-88404  &   114.851067     &  -88.6766875     &  00036747002 &  2007-12-11 11:18:10  &  2007-12-11 22:45:57  \\
  1SWXRT J072442.5-883748  &  1RXSJ073856.5-88404  &   111.177442     &  -88.6301269     &  00036747002 &  2007-12-11 11:18:10  &  2007-12-11 22:45:57  \\
  1SWXRT J125106.6-884154  &  1RXSJ125047.2-88415  &   192.777875     &  -88.6986097     &  00036735001 &  2008-03-15 08:19:37  &  2008-03-15 11:42:58  \\
  1SWXRT J123848.5-884129  &  1RXSJ125047.2-88415  &   189.702258     &  -88.6914839     &  00036735001 &  2008-03-15 08:19:37  &  2008-03-15 11:42:58  \\
  1SWXRT J000152.2-870707  &  CRATESJ0011-8706     &  0.467620800     &  -87.1187878     &  00039230001 &  2009-08-20 13:47:50  &  2009-08-20 23:35:56  \\
  1SWXRT J000851.6-870657  &  CRATESJ0011-8706     &   2.21538750     &  -87.1158667     &  00039230001 &  2009-08-20 13:47:50  &  2009-08-20 23:35:56  \\
  1SWXRT J001152.0-870624  &  CRATESJ0011-8706     &   2.96666670     &  -87.1069058     &  00039230001 &  2009-08-20 13:47:50  &  2009-08-20 23:35:56  \\
  1SWXRT J002143.9-865701  &  CRATESJ0011-8706     &   5.43310000     &  -86.9504700     &  00039230001 &  2009-08-20 13:47:50  &  2009-08-20 23:35:56  \\
  \hline
\end{tabular}
\begin{tabular}{|cccccc|}
  \hline
  \hline
   F\_RATE          &  F\_RATE\_ERR     &       F\_FLUX      &      F\_FLUX\_ERR  &    F\_PROB        &      F\_SNR                \\
  \hline
   0.29400000349E-02 &   0.10000000475E-02 &   0.12941880098E-12 &   0.44020000725E-13 &  0.16440000472E-08 &    2.8469998837      \\
   0.67446837202E-02 &   -99.000000000     &   0.29690098578E-12 &   -99.000000000     &   0.0000000000     &    0.0000000000      \\
   0.45307222754E-02 &   -99.000000000     &   0.19944239131E-12 &   -99.000000000     &   0.0000000000     &    0.0000000000      \\
   0.42099999264E-02 &   0.12000000570E-02 &   0.17362039540E-12 &   0.49488004975E-13 &  0.13740000619E-12 &    3.5369999409      \\
   0.30799999833E-01 &   0.44999998063E-02 &   0.10561320698E-11 &   0.15430499489E-12 &   0.0000000000     &    6.7829999924      \\
   0.11199999601E-01 &   0.28999999631E-02 &   0.43668795793E-12 &   0.11307099734E-12 &   0.0000000000     &    3.8340001106      \\
   0.88299997151E-02 &   0.15999999596E-02 &   0.36414919474E-12 &   0.65984002116E-13 &   0.0000000000     &    5.3550000191      \\
   0.12900000438E-01 &   0.20000000950E-02 &   0.69131104920E-12 &   0.10718000516E-12 &   0.0000000000     &    6.5789999962      \\
   0.99999997765E-02 &   0.17000000225E-02 &   0.50119999297E-12 &   0.85203999617E-13 &   0.0000000000     &    5.8779997826      \\
   0.39300001226E-02 &   0.12000000570E-02 &   0.16207320859E-12 &   0.49488004975E-13 &  0.74759998539E-11 &    3.3729999065      \\
  \hline
\end{tabular}
\begin{tabular}{|cccccc|}
  \hline
  \hline
    S\_RATE         &  S\_RATE\_ERR     &        S\_FLUX      &     S\_FLUX\_ERR  &        S\_PROB    &   S\_SNR                  \\
  \hline
   0.44355490245E-02 &   -99.000000000     &   0.79351970958E-13 &   -99.000000000     &   0.0000000000     &    0.0000000000      \\
   0.38999998942E-02 &   0.12000000570E-02 &   0.69770998332E-13 &   0.21468000920E-13 &  0.14899999637E-11 &    3.2990000248      \\
   0.25700000115E-02 &   0.91000000248E-03 &   0.45977300743E-13 &   0.16279900401E-13 &  0.43620000945E-08 &    2.8289999962      \\
   0.30499999411E-02 &   0.10000000475E-02 &   0.53984997760E-13 &   0.17700000265E-13 &  0.31590001448E-10 &    3.0299999714      \\
   0.24100000039E-01 &   0.38999998942E-02 &   0.41789399090E-12 &   0.67625992425E-13 &   0.0000000000     &    6.1009998322      \\
   0.75699998997E-02 &   0.24000001140E-02 &   0.13156660548E-12 &   0.41712001962E-13 &  0.19780000120E-11 &    3.1840000153      \\
   0.74200001545E-02 &   0.15000000130E-02 &   0.13133400346E-12 &   0.26549998704E-13 &   0.0000000000     &    5.0000000000      \\
   0.96899997443E-02 &   0.17000000225E-02 &   0.16744319418E-12 &   0.29376000466E-13 &   0.0000000000     &    5.7859997749      \\
   0.72599998675E-02 &   0.13999999501E-02 &   0.12646918831E-12 &   0.24387997928E-13 &   0.0000000000     &    5.0960001945      \\
   0.35900000948E-02 &   0.10999999940E-02 &   0.63543002240E-13 &   0.19469999275E-13 &  0.41699998489E-11 &    3.2839999199      \\
  \hline
\end{tabular}
\begin{tabular}{|cccccc|}
  \hline
  \hline
    H\_RATE         &     H\_RATE\_ERR  &      H\_FLUX        &      H\_FLUX\_ERR &      H\_PROB       &      H\_SNR               \\
   \hline            
   \hline
   0.26165884919E-02 &   -99.000000000     &   0.22204369691E-12 &   -99.000000000     &    0.0000000000     &   0.0000000000      \\
   0.93410944100E-03 &   -99.000000000     &   0.79268528048E-13 &   -99.000000000     &    0.0000000000     &   0.0000000000      \\
   0.79634931171E-03 &   -99.000000000     &   0.67578206215E-13 &   -99.000000000     &    0.0000000000     &   0.0000000000      \\
   0.35612971988E-02 &   -99.000000000     &   0.30121453834E-12 &   -99.000000000     &    0.0000000000     &   0.0000000000      \\
   0.70600002073E-02 &   0.20999999251E-02 &   0.56239962774E-12 &   0.16728599842E-12 &   0.18099999716E-13 &   3.3510000706      \\
   0.87090013549E-02 &   -99.000000000     &   0.73503974175E-12 &   -99.000000000     &    0.0000000000     &   0.0000000000      \\
   0.27999999002E-02 &   0.92000002041E-03 &   0.23682398815E-12 &   0.77813603271E-13 &   0.18029999369E-10 &   3.0369999409      \\
   0.55800001137E-02 &   0.13000000035E-02 &   0.53484300322E-12 &   0.12460499754E-12 &    0.0000000000     &   4.1799998283      \\
   0.37000000011E-02 &   0.10000000475E-02 &   0.34335999755E-12 &   0.92800001353E-13 &   0.44410000952E-15 &   3.6170001030      \\
   0.24837893434E-02 &   -99.000000000     &   0.21007891607E-12 &   -99.000000000     &    0.0000000000     &   0.0000000000      \\
  \hline
\end{tabular}
\begin{tabular}{|ccccccc|}
  \hline
  \hline
       EXPOSURE   &         HR           &         HR\_ERR    &  POS\_ERR        &        N$_H$  &      BETA      &    BETA\_ERR   \\    
   \hline            
   \hline
    3906.6989746  &     99.000000000     &    99.000000000    & 5.8758726120     &   0.11026673845E+22 &   99.000000000     &    0.0000000000     \\
    3906.6989746  &     99.000000000     &    99.000000000    & 5.3863286972     &   0.10955931443E+22 &   99.000000000     &    0.0000000000     \\
    4360.2348633  &     99.000000000     &    99.000000000    & 6.1900162697     &   0.10954888578E+22 &   99.000000000     &    0.0000000000     \\
    4360.2348633  &     99.000000000     &    99.000000000    & 5.4950857162     &   0.10945900378E+22 &   99.000000000     &    0.0000000000     \\
    2059.5390625  &   -0.54685002565     &   0.11868000031    & 3.2425003052     &   0.69874664114E+21 &   1.0235582590     &   0.30429711938     \\
    2059.5390625  &     99.000000000     &    99.000000000    & 5.0195226669     &   0.70344396592E+21 &   99.000000000     &    0.0000000000     \\
    4825.4760742  &   -0.45205000043     &   0.15346999466    & 3.7346503735     &   0.88651529077E+21 &  0.85407936573     &   0.37225008011     \\
    4825.4760742  &   -0.26916000247     &   0.13526000082    & 3.5496118069     &   0.88294752508E+21 &  0.46099326015     &   0.28463935852     \\
    4825.4760742  &   -0.32482001185     &   0.14848999679    & 3.5868091583     &   0.88570900570E+21 &  0.57517987490     &   0.32556995749     \\
    4825.4760742  &     99.000000000     &    99.000000000    & 5.4949035645     &   0.88218212425E+21 &   99.000000000     &    0.0000000000     \\
  \hline
\end{tabular}
\end{center}
\label{cata}
\end{table*}


\subsection{Uncertainties and source reliability}

{\it Detect} count rates are associated with their statistical
(Poissonian) uncertainties. These errors are propagated to the flux
estimates, but here the main uncertainty is the variety of the
spectral behaviour of different sources. In order to determine the
flux variation with the spectral parameters, we estimate the count
rate-to-flux conversion factors for a wide range of spectral slopes
($\Gamma = 0 -2$) and Hydrogen column densities ($N_H = 10^{19} -
10^{22}$ cm$^{-2}$). The conversion factors are in the range $(2.9 -
15)\times 10^{-11}$, $(0.9-1.5) \times 10^{-11}$ and $(8.1-17)\times
10^{-11}$ erg cm$^{-2}$ s$^{-1}$\ for the F, S and H band,
respectively. The conversion factor for the F band is more sensitive
to the spectral shape than for the S and H bands, because this band is
wider.

Concerning the source positions,
their errors are both statistical and systematic, with the total positional
uncertainty being:

$$\sigma_{pos}= \sqrt{\sigma^2_{stat}+\sigma^2_{sys}}.\eqno (2)$$

The systematic error $\sigma_{sys}$ is due to the uncertainty on the
XRT aspect solution. This quantity has been estimated by Puccetti et
al. (2011) by cross-correlating a sub-sample of bright sources of
their XRT-deep catalog with the SDSS optical galaxy catalog. They
found that the mean $\sigma_{sys}$ at the $68\%$ confidence level is
$2.05$ arcsec, a value consistent with previous results by Moretti et
al. (2006). This value represents the number we will adopt in
estimating the positional error in Eq. 2. The statistical variance
$\sigma^2_{stat}$ is instead inversely proportional to the source number
counts. 

To assess the reliability of our detections we must address the
possibility of source confusion.
The source confusion issue arises when two close sources are detected
as a single one. This problem may be important if the distances
between two objects is of the order of the cell detection of the
algorithm {\it detect}. To evaluate the possibility of source
confusion, we compute the probability of finding two sources with a
X-ray flux higher than a certain threshold $F_{lim}$, lying at a
distance smaller than $\theta_{min}$:

$$ P(<\theta_{min}) = 1-e^{-\pi\;N\;\theta^2_{min}}. \eqno (3)$$

Here we adopt as $\theta_{min}$ twice the typical size of the cell
detection box ($4$ pixels or $9.44$ arcsec), while $N$ is the number
counts corresponding to $F_{lim}$, which can be evaluated, e.g., from
the C-COSMOS survey (Elvis et al. 2009). Our deepest field has an
exposure of $\sim 100$ ks. Using the $F_{lim}$ corresponding to the
count rates of the faintest sources detected in this field ($\sim
1.7\times 10^{-4}$ and $\sim 1.5\times 10^{-4}$ cts/s in the S and H
band, respectively), we find that the source confusion probability is
less than $3\%$ in both the S and H band. This is of course the field
in which the source confusion probability is highest. For fields of
$\sim 10$ ks ($\sim 93 \%$ of our sample has exposures $<10$ ks) the
flux limits are shallower by a factor of $\sim 3$. Applying Eq. (3) to
these fields results in a probability of source confusion of $\sim
0.9\%$ and $\sim 0.3\%$ in the S and H band, respectively. This means
that source confusion is negligible in our sample.

\subsection{1SWXRT description}

The final catalog comprises $32$ field parameters for each
entry. Source name, position, count rates and fluxes, exposure,
hardness ratio and galactic $N_H$ are reported, together with the
corresponding uncertainties and/or reliabilities. A full description
of all the parameters is presented in Table 3. Table 4 gives instead
the first ten entries of the catalog as an example. 

\section{Scientific use of the catalog}

A full exploitation of the scientific data presented in this work is
far beyond the scope of the present paper. Nevertheless, we would like
to draw the reader's attention to some of the scientific topics that
can be addressed using 1SWXRT.

\subsection{Short-term variability}

As stated in the previous section, in our analysis we do not merge
observations pointing to the same field, so we can study the
variability of sources observed more than once. Since many
observations are often performed consecutively, this enables to
determine short-term variability for the involved sources. 

Our database comprises 12,908 sources observed at least twice. Among
these, we select all sources detected in each observation in the soft
or hard band. 7,936 and 2,113 sources are detected in the soft and
hard band, respectively. Fig. 9 plots the distribution of the number
of sources observed many times, while Fig. 10 displays the histogram
for the variability as a function of the $\sigma$ significance. The
number of sources in the soft band with a variation larger than $3
\sigma$ and $5 \sigma$ is 1,774 and 623, respectively, i.e., a
fraction of $22$\% and $7.7$\% of the total soft sources. Similarly,
the number of sources in the hard band with a variation larger than $3
\sigma$ and $5 \sigma$ is $447$ and $148$, respectively, i.e., a
fraction of $23$\% and $7.6$\% of the total hard sources. Thus,
variability is observed in both bands, and with similar
trends. However, some tens of sources show extreme variability
(Fig. 10). The ratio of such extreme variable sources
with respect to the total number grows stronger in the hard band with
respect to the soft one as the significance of the variability
increases. For example, the fraction of sources which vary at more
than $10 \sigma$ is $1.7$\% and $1.9$\% in the soft and hard band,
respectively, while at the $20 \sigma$ level, the fractions become
$0.4$\% (soft) and $0.7$\% (hard).

We then select all the sources observed at least 5 times. Fig. 11
shows the cumulative distribution of the statistical significance of
the variability for sources with five observations or more. This
variability significance has been computed with respect both to the
maximum and to the minimum fluxes. It interesting to note that the
variability is more pronounced when considering the maximum fluxes. In
other words, the average fluxes are in general closer to the minimum
values than to the maximum ones. This could be a possible indication
that we are observing short-duration flares in some sources, with the
normal state being close to the minimum value observed.

\begin{figure}
\centering
\includegraphics[height=8.4cm,width=9.5cm,angle=-0]{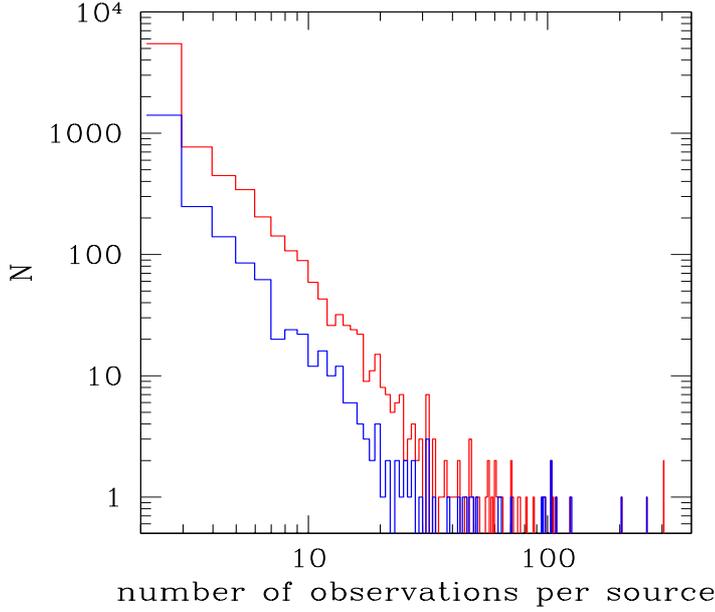}
\caption{The distribution of the number of sources observed more than
  once. Red line refers to the soft band, blue line to the hard one.}
\end{figure}

\begin{figure}
\centering
\includegraphics[height=8.4cm,width=9.5cm,angle=-0]{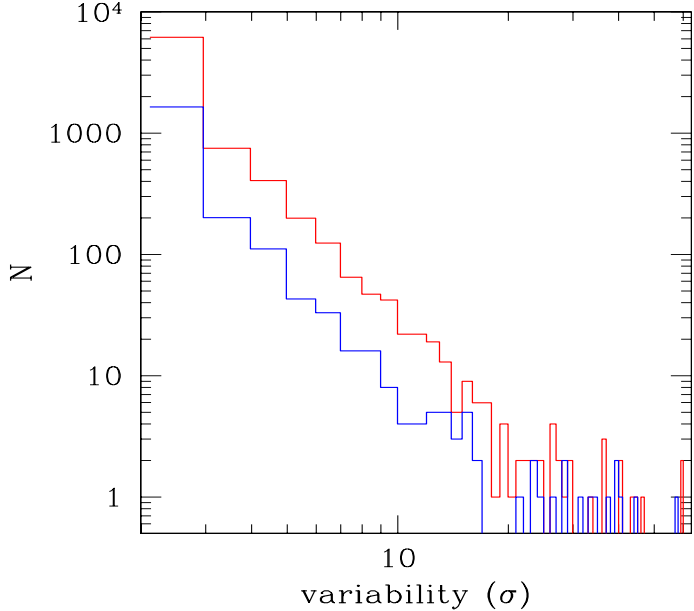}
\caption{The distribution of the source variability expressed as a
  function of the $\sigma$ significance. Red line refers to the soft
  band, blue line to the hard one.}
\end{figure}

\begin{figure}
\centering
\includegraphics[height=8.4cm,width=9.5cm,angle=-0]{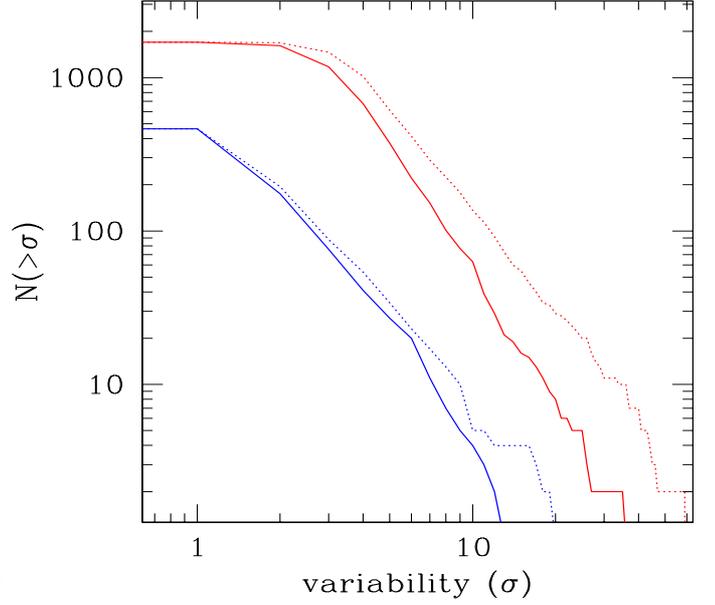}
\caption{ The cumulative distribution of the statistical significance
  of the variability for sources with five observations or more. Red
  (Blue) lines refer to the soft (hard) band. Solid (dotted) lines
  refer to the variability significance of the minimum (maximum) flux
  values with respect to the average ones.}

\end{figure}



\subsection{Soft sources}

We can use our dataset to study sources showing emission in the soft
band only. Among these, one important class is represented by isolated
neutron stars (INS, see, e.g., Treves et al. 2000; Haberl et al. 2003;
Haberl 2004).

INS are blank field sources, i.e., X-ray sources with no or very faint
counterparts in other wavelength domains. Concerning the
X-ray-to-optical flux ratio, values of $f_X/f_{opt} > 10^{3}$ define
the INS class, but in some cases values as high as $10^5$ have been
reported. The X-ray emission is supposed to be produced by some
residual internal energy (coolers) or because they are interacting
with the interstellar medium (accretors). The INS X-ray spectrum is
well fitted by a soft blackbody, with temperatures of $\sim 100$
eV. This means that basically no X-ray emission above $\sim 2$ keV is
expected. Given the low column densities measured for these objects,
the emission is consistent with being produced from the neutron star
surface (see, e.g., Walter \& Lattimer 2002). Other characteristics
often exhibited by these sources (coolers) are a periodicity of $\sim
5 - 10$ s, absorption features below $1$ keV and closeness. These
elusive sources are of extreme importance, because they could
represent $\sim1\%$ of the total number of stars in our Galaxy. To
pinpoint their properties means to understand the end-point of the
evolution of a large class of stars. To date, only $8-10$ objects of
this class have been identified.

In order to check our catalog for the presence of INS, and in general
to categorize the soft objects, we selected all sources that do not
show emission in the full and hard band. When considering objects
observed more than once, we excluded from our analysis all sources in
which there is a detection in the full or hard band in at least one
observation. This helps us to include in our sample just genuine soft
emitters, and to exclude part of the sources that are possibly not
detected in the hard band due to low exposure times.

We selected $2087$ objects following the above criteria. Fig. 12
displays the $0.5-2$ keV flux distribution for these sources. The
histogram bin size is set to $0.05$ dex. Since the soft band is in
general more sensitive than the hard one, the faint part of this
distribution can still comprise normal sources that are not detected
in the $2-10$ keV range due to a flux level below the sensitivity
threshold. However, we determined the number of XRT sources featuring
at least $50$ or $100$ counts in the $0.5-2$ keV band, without
detection in the $2-10$ one. We obtain $7$ sources with at least $50$
counts. Of these, $3$ have more than $100$ counts.  These seven
objects are good INS candidates.



\begin{figure}
\centering
\includegraphics[height=8.4cm,width=9.5cm,angle=-0]{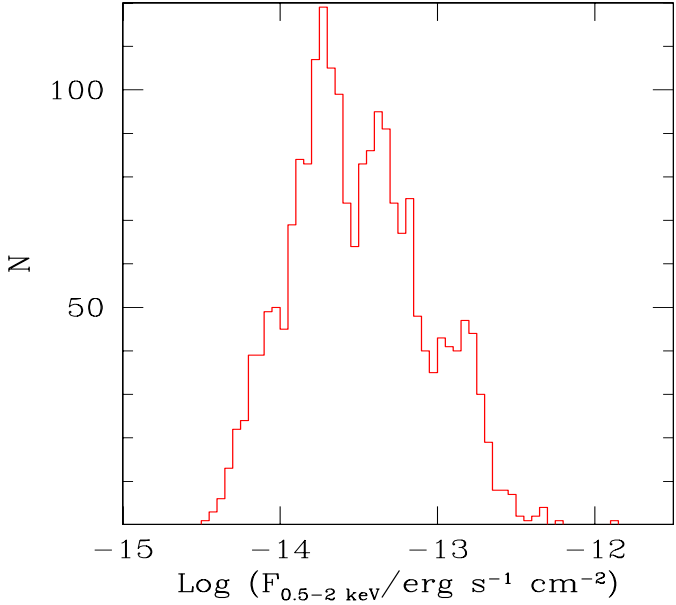}
\caption{The distribution of the $0.5-2$ keV flux for the sources
  detected in the soft band only.}
\end{figure}

\subsection{Hard sources}

In a similar way to what described in the previous sub-section, we can
search our dataset for sources which show emission in the hard band
only. To categorize the hard objects, we selected all sources that do
not show emission in the full and soft band. When considering objects
observed more than once, we excluded from our analysis all sources in
which there is a detection in the full or soft band in at least one
observation. This helps us to include in our sample just genuine hard
emitters, and to exclude part of the sources that are possibly not
detected in the soft band due to a low exposure time coupled with an
unusual background level. $308$ objects in our dataset fulfill the
above criteria. Fig. 13 displays the $2-10$ keV flux distribution for
these sources. The histogram bin dimension is set to $0.1$ dex. The
hard band is less sensitive than the soft one.  Thus, contrary to the
case of the soft sources, we are confident that this sub-sample
contains genuine hard sources only.

The main type of objects contributing to this sub-sample are expected
to be obscured Active Galactic Nuclei (AGN), whose discovery and study
is very important both to study the properties and evolution of the
accretion process onto supermassive black holes residing at the center
of galaxies and to determine their contribution to the X-ray
background, in particular to its peak emission in the 20-30 keV band
that still remains largely unexplained (see, e.g., Gilli et al. 2007;
Treister et al. 2009, and references therein).  In future works we
will investigate on the nature of these sources in order to determine
their properties and nature.

\begin{figure}
\centering
\includegraphics[height=8.4cm,width=9.5cm,angle=-0]{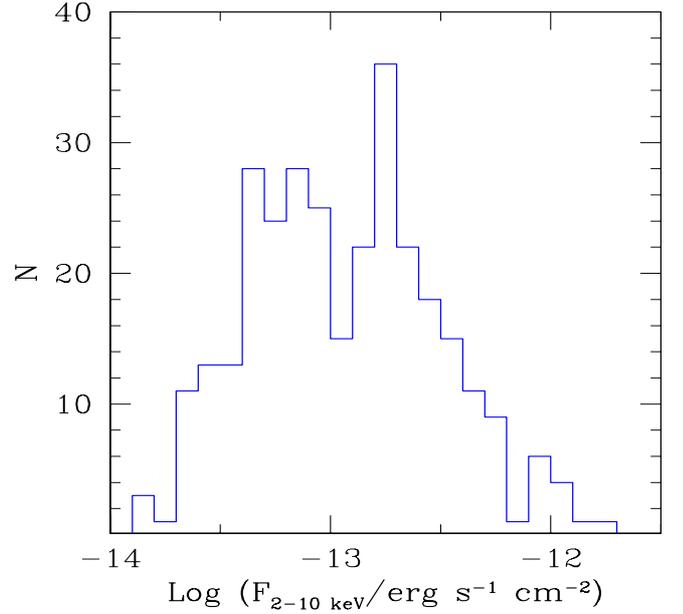}
\caption{The distribution of the $2-10$ keV flux for the sources
  detected in the hard band only.}
\end{figure}

\subsection{Cross-correlation with multi-wavelength catalogs}

Our catalog can be cross-correlated with multi-wavelength ones, to
obtain statistical information about specific class of sources. Here,
we cross-correlated the XRT catalog with BZCAT, a multifrequency
catalogue of blazars (Massaro et al. 2009). We stress that this is
just an example, and that many more cross-correlations with other
catalogs can be performed to fully exploit 1SWXRT.

Blazars are radio loud AGN pointing their jets in the direction of the
observer (see e.g. Urry \& Padovani 1995). They come in two main
subclasses, the Flat Spectrum Radio Quasars (FSRQs), which show
strong, broad emission lines in their optical spectrum, just like
radio quiet QSOs, and BL Lacs, which are instead characterized by an
optical spectrum, which at most shows weak emission lines or is
completely featureless. The strong non-thermal radiation is composed
of two basic parts forming two broad humps in the $\nu$ vs. $\nu
F_{\nu}$ plane, the low-energy one attributed to synchrotron
radiation, and the high-energy one, usually thought to be due to
inverse Compton radiation (Abdo et al. 2010). The peak of the
synchrotron hump (\nup) can occur at different frequencies. In FSRQs
\nup~never reaches very high values (\nup~$\lsim 10^{14.5}$ Hz),
whereas the \nup~of BL Lacs can reach values as high as \nup~$\gsim
10^{18}$ Hz (e.g. Giommi et al. 2012).

The cross-correlation between the BZCAT and 1SWXRT catalogs has been
performed by matching the coordinates over an error radius of $0.2$
arcmins. We found $938$ sources in 1SWXRT with a BZCAT counterpart. Of
these, $524$ are FSRQs and $414$ are BL Lacs.  Fig. 14 shows the X-ray
spectral index distribution for these sources. It is evident that BL
Lac distribution is softer than FSRQ one. This is because the X-ray
$0.5 - 10$ keV band samples on average the high energy tail of the
synchrotron emission in BL Lacs, where $\nu F_{\nu}$ is decreasing. On
the other hand, the same energy band describes, on average, the low
energy tail of the inverse Compton emission in FSRQs, where $\nu
F_{\nu}$ is instead increasing.  For comparison, Fig. 14 plots also
the X-ray spectral index of the stars, obtained by cross-correlating
the XRT catalog with the Smithsonian Astrophysical Observatory Star
Catalog (SAO), and that of the unidentified sources.

\begin{figure}
\centering
\includegraphics[height=8.4cm,width=9.5cm,angle=-90]{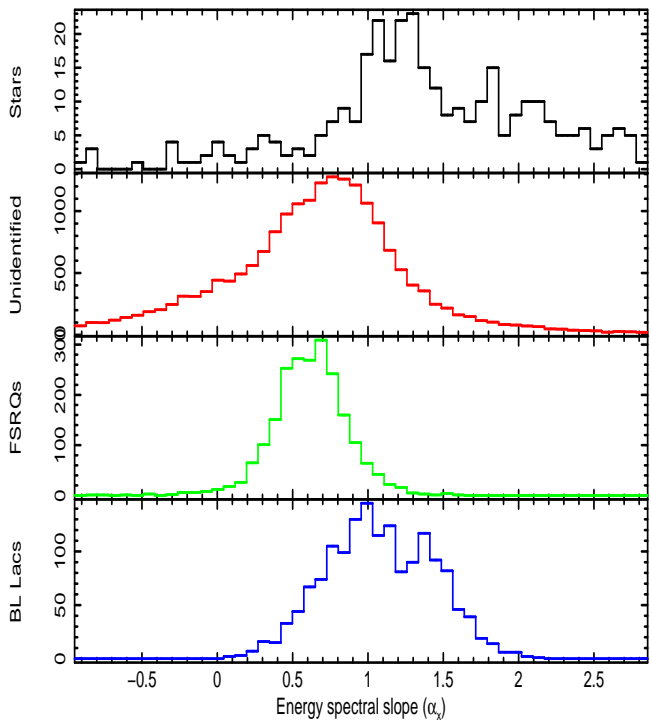}
\caption{The distribution of the X-ray spectral index for specific
  source types in our catalog.}
\end{figure}

\section{Summary and conclusions}

We have reduced and analyzed all the observations performed by {\it
  Swift}-XRT in PC mode with an exposure time longer than $500$ s,
during its first seven years of operations, i.e., between 2005 and
2011. Approximately 35,000 XRT fields have been analyzed, with net
exposures (after screening and filtering criteria being applied)
ranging from $500$ s to $100$ ks. The total, net exposure time is
$\sim 140$ Ms.

The purpose of this work was to create a catalog (1SWXRT) of 
all the point like sources detected in these observations. To this
purpose, we run the XIMAGE {\it detect} algorithm to all our fields,
and then removed spurious and extended sources through visual
inspection of the XRT observations. The total number of point-like
objects detected is $89,053$, of which $2,166$ are GRB detections (so
transient sources by definition) and $1,947$ are sources affected by
pile-up. Thus, our final version of the catalog comprises $84,992$
entries, which define the ``good'' sample. Many entries represent the
same sources, since several portions of the sky have been observed
many times by XRT. To estimate an approximate number of distinct,
celestial sources, we compress our catalog over a radius of $12$
arcsec, a typical positional uncertainty value in faint XRT
sources. In other words, all entries closer than $12$ arcsec are
counted once, and the result of this procedure is $\sim$ 36,000
distinct sources.

For all the entries of 1SWXRT, we determined the position, the
detection probability and the signal-to-noise ratio. Count rates were
estimated in the $0.3-10$, $0.3-3$ and $2-10$ keV bands. Each source
has a detection in at least one of these bands, with $\sim$ 80,000,
$\sim$ 70,000 and $\sim$ 25,500 sources detected in the full, soft and
hard band, respectively. $90\%$ upper limits were provided in case of
missing detection in one or two of these bands. The count rates were
converted into fluxes in the $0.5-10$, $0.5-2$ and $2-10$ keV X-ray
bands. The flux interval sampled by the detected sources is $7.4\times
10^{-15} - 9.1\times 10^{-11}$, $3.1\times 10^{-15} - 1.1\times
10^{-11}$ and $1.3\times 10^{-14} - 9.1\times 10^{-11}$ erg cm$^{-2}$
s$^{-1}$ for the full, soft and hard band, respectively. Among the
possible scientific uses of 1SWXRT, we discussed the possibility to
study short-term variability, the identification of sources emitting
in the soft or hard band only, and the cross correlation of our
catalogue to multi-wavelength ones.

\begin{acknowledgements}
  We thank the referee for a quick and careful reading of the
  manuscript.  This work has been supported by ASI grant I/004/11/0.
  JPO acknowledges financial support from the UK Space Agency
\end{acknowledgements}


\begin{thebibliography}{DUM}

\bibitem[]{} Abbey T., Carpenter J., Read A. et al. 2006, The X-Ray Universe 2005, 604, 943
\bibitem[]{} Abdo A.A., Ackermann M., Agudo I. et al. 2010, 2010, ApJ, 716, 30,
\bibitem[]{} Bevington P.R. \& Robinson K. 1992, Data Reduction and
  Error Analysis for the Physical Sciences (the McGraw-Hill Companies,
  Inc.)
\bibitem[]{} Barthelmy S.D., Barbier L.M.,  Cummings, J. R. et al. 2005, SSR, 120, 143
\bibitem[]{} Burrows D.N., Hill J.E., Nousek J.A. et al. 2005, SSR 120, 165
\bibitem[]{} Capalbi M., Perri M. Saija B. Tamburelli F. \& Angelini L. 2005, http://heasarc.nasa.gov/docs/swift/analysis/xrt swguide v1 2.pdf
\bibitem[]{} Elvis M., Civano F., Vignali C., et al. 2009, ApJS, 184, 158
\bibitem[]{} Evans I.N., Primini F.A., Glotfelthy K.J et al. 2010, ApJ, 189, 37

  437, 845
\bibitem[]{} Gehrels N., Chincarini G., Giommi P., et al. 2004, ApJ 621, 558
\bibitem[]{} Gilli, R., Comastri, A., Hasinger, G. 2007, A\&A, 463, 79
\bibitem[]{} Giommi P., et al., 2012, A\&A, 514, 160

\bibitem[]{} Haberl F, Schwope A.D., Hambaryan V., Hasinger G. \& Motch C. 2003, A\&A, 406, 471
\bibitem[]{} Haberl F 2004, MemSAIt, 75, 454
\bibitem[]{} Hill J.E., Burrows D.N., Nousek J.A. et al. 2004, SPIE, 5165, 217
\bibitem[]{} Kalberla P.M.W., Burton W.B., Hartmann D. et al. 2005, A\&A, 440, 775
\bibitem[]{} Massaro E, Giommi P., Leto C., Marchegiani P., Maselli
  A., Perri M., Piranomonte S., Sclavi S. 2009, A\&A, 495, 691
\bibitem[]{} Moretti A, Perri M., Capalbi M., et al. 2006, A\&A, 448, L9 
\bibitem[]{} Puccetti S., Capalbi M., Giommi P. et al. 2011, A\&A,
  528, 122
\bibitem[]{} Romano P 2012, Mem SAIt,  19, 306
\bibitem[]{} Sari R. \& Piran T. 1998, MNRAS, 287, 110
\bibitem[]{} Tundo E., Moretti A., Tozzi P., Teng L., Rosati P., Tagliaferri G., Campana S. 2012, A\&A, 547, 57
\bibitem[]{} Urry M. \& Padovani P. 1995, PASP, 107, 83
  Nature 461, 1254
\bibitem[]{} Treister E., Urry C.M., Virani S. 2009, ApJ, 696, 110
\bibitem[]{} Treves A., Turolla R., Zane S. \& Colpi M. 2000, PASP, 112, 297
  468, 83
\bibitem[]{} Walter F.M. \& Lattimer J.M. 2002, ApJ, 576, 145
\bibitem[]{} Watson M.G., Schr\"oder A.C., Fyfe D. et al. 2009, A\&A,
  493, 339


\end{thebibliography}
\end{document}